\documentclass[11pt, cls, onecolumn]{IEEEtran}

\usepackage{subfigure,cite,graphicx,epsfig,dblfloatfix}
\usepackage{amsmath,amssymb,mathrsfs}
\usepackage{psfrag}
\usepackage{multirow}
\usepackage{tabularx}

\let\oldbrace\{
\def\{{\oldbrace\kern0.5pt}

\let\P\relax
\DeclareMathOperator\P{\sf P}

\newtheorem{theorem}{Theorem}
\newtheorem{example}{Example}
\newtheorem{remark}{Remark}
\newtheorem{lemma}{Lemma}
\newtheorem{proposition}{Proposition}
\newtheorem{corollary}{Corollary}
\newtheorem{definition}{Definition}

\begin{document}
\title{Interactive Computation of Type-Threshold Functions in Collocated Broadcast--Superposition Networks}
\author{Chien-Yi Wang\IEEEmembership{, Student Member, IEEE}, Sang-Woon Jeon\IEEEmembership{, Member, IEEE}, and\\ Michael Gastpar\IEEEmembership{, Member, IEEE}
\thanks{This work was supported in part by the European ERC Starting Grant 259530-ComCom. The second author was also funded in part by the MSIP (Ministry of Science, ICT \& Future Planning), Korea in the ICT R \& D Program 2013. The material in this paper was presented in part at the IEEE International Symposium on Information Theory (ISIT), Turkey, Istanbul, July 2013.}
\thanks{C.-Y. Wang is with the School of Computer and Communication Sciences, Ecole Polytechnique F{\'e}d{\'e}rale de Lausanne (EPFL), Lausanne,
Switzerland (e-mail: chien-yi.wang@epfl.ch).}
\thanks{S.-W. Jeon is with the Department of Information and Communication Engineering, Andong National University, South Korea (e-mail: swjeon@anu.ac.kr).}

\thanks{M. Gastpar is with the School of Computer and Communication Sciences, Ecole Polytechnique F{\'e}d{\'e}rale de Lausanne (EPFL), Lausanne, Switzerland and the Department of Electrical Engineering and Computer Sciences, University of California, Berkeley, CA, USA (e-mail: michael.gastpar@epfl.ch).}
}
\maketitle

\begin{abstract}
In wireless sensor networks, various applications involve learning one or multiple functions of the measurements
observed by sensors, rather than the measurements themselves. This paper focuses on type-threshold functions, e.g., the maximum and indicator functions. Previous work studied this problem under the collocated collision network model and showed that under many probabilistic models for the measurements, the achievable computation rates converge to zero as the number of sensors increases.
This paper considers two network models reflecting both the broadcast and superposition properties of wireless channels: the collocated linear finite field network and the collocated Gaussian network. A general multi-round coding scheme exploiting not only the broadcast property but particularly also the superposition property of the networks is developed. Through careful scheduling of concurrent transmissions to reduce redundancy, it is shown that given any independent measurement distribution, all type-threshold functions can be computed reliably with a non-vanishing rate in the collocated Gaussian network, even if the number of sensors tends to infinity.
\end{abstract}


\section{Introduction} 
To date, wireless sensor networks have been deployed for various applications. Typically, a sensor network consists of 
a single fusion center and multiple sensors measuring certain parameters. Sensor deployment can be costly, so the lifetime of sensors is expected to be months or even years. Therefore, {\it power efficiency} becomes an important issue for system design. 
Traditionally, sensors simply convey all the measured parameters to the fusion center. However, for many applications, the fusion center is only interested in acquiring an {\it indication} or, more generally, a {\it function} of the parameters, rather than the parameters themselves. For example, in forest fire detection, only an alarm signal is needed instead of the whole temperature and/or humidity readings.

In this paper, we assume that the fusion center wants to collect multiple instances of the same function and the sensors are allowed to code over long sequences of measurements. The performance metric considered in this paper is {\it computation rate}, i.e., the number of functions computed reliably per channel use. The problem of function computation in wireless sensor networks has recently received significant attention. One interesting formulation was developed by Giridhar and Kumar \cite{Giridhar:05}. First, they assumed that all nodes are collocated, which means any transmit signal is received by all nodes except the sender. Second, they modeled the wireless medium as a collision channel, i.e., concurrent transmissions by multiple nodes result in collisions. They considered the class of {\em symmetric} functions and particularly the subclasses of {\em type-sensitive} and {\em type-threshold} functions. The main focus of this paper is the class of type-threshold functions which includes the maximum, minimum, and indicator functions as special cases. Intuitively, type-threshold functions have {\it relatively small} ranges.

For the computation of type-threshold functions under the collocated collision network model, Giridhar and Kumar showed that the worst-case scaling of computation rate with respect to the number of sensors $M$ is $\Theta(\frac{1}{\log M})$. Here the worst case means the worst source (or measurement) distribution for computing the desired function, which may depend on $M$.
Later, Ma, Ishwar, and Gupta \cite{Ma:12} followed the same model and studied the problem within the framework of interactive source coding. Still, the worst-case scaling of computation rate for type-threshold functions is $\Theta(\frac{1}{\log M})$.
On the other hand, Kowshik and Kumar \cite{Kowshik:09} showed that, if the source distribution is independent of $M$, then the computation rate $\Theta(1)$ is achievable. Furthermore, Subramanian, Gupta, and Shakkottai \cite{Subramanian:07} showed that the computation rate $\Theta(1)$ is achievable if the number of nodes within a direct communication range is upper bounded by a fixed number independent of $M$. 

To study the fundamentals of type-threshold function computation in wireless networks, we consider two network models reflecting both the broadcast and superposition properties of wireless channels: the collocated linear finite field network and the collocated Gaussian network. We propose a novel coding scheme termed {\it multi-round group broadcast}, which is an extension of type computation coding \cite{JeonWangGastparISIT:13} to the framework of interactive computation. We show that, for any independent source distribution, all type-threshold functions are reliably computable with a non-vanishing rate in the collocated Gaussian network, even if the number of sensors tends to infinity. Whereas previous work inherently assumes that sending multiple signals causes collisions and only exploit the broadcast property of wireless channels to achieve the computation rate $\Theta(\frac{1}{\log M})$, our result shows that in general, exploiting both the broadcast and superposition properties is necessary to achieve the computation rate $\Theta(1)$. Table I summarizes the achievable scaling laws for collocated networks. 

An outline of the paper is as follows. In Section \ref{sec:prob_state}, we provide our problem formulation defining network models and type-threshold functions. In Section \ref{sec:IRR}, as a preliminary, we extend the existing schemes for collocated collision networks to collocated broadcast--superposition networks. In Section \ref{sec:entropy}, we introduce a set of auxiliary random variables, also termed {\it descriptions} in this paper, with an analysis on its entropy. These descriptions serve as the building blocks of the proposed multi-round group broadcast which is introduced in Section \ref{sec:MRGB}. In particular, Section \ref{sec:Fp} and Section \ref{sec:Gaussian} are devoted to the collocated linear finite field network and the collocated Gaussian network, respectively. A simple cut-set based upper bound is given in Section \ref{sec:cut-set}. Finally, we conclude in Section \ref{sec:conclude}.

\begin{table}[t!]
\scriptsize
\label{table:summary}
\caption{Achievable scaling law for the number of sensors}
\begin{center}
\begin{tabular}{|c|cc|cc|} 
\hline 
& \multicolumn{2}{ c }{Collocated collision networks}   & \multicolumn{2}{ |c| }{Collocated Gaussian networks} \\
\hline
Full data & \hspace{1cm}$\Theta\left(\frac{1}{M}\right)$ &  & \hspace{1.7cm}$\Theta\left(\frac{1}{M}\right)$ & \\ \hline
Symmetric functions & \hspace{1cm}$\Theta\left(\frac{1}{M}\right)$ & \cite{Giridhar:05} & \hspace{1.7cm} $\Theta\left(\frac{1}{\log M}\right)$ &  \cite{JeonWangGastparISIT:13}\\ \hline
Type-threshold functions & \hspace{1cm} $\Theta\left(\frac{1}{\log M}\right)$ & \cite{Giridhar:05} & \hspace{1.7cm} $\Theta\left(1\right)$ & (this work)\\ \hline
\end{tabular}
\end{center}
\end{table}

{\it Notation:}
Denote by $(\mathbb{R},+,\times)$ the field of real numbers and by $(\mathbb{F}_{p},\oplus,\otimes)$ the finite field of order $p$, where $p$ is assumed to be prime in this paper. Also, we denote $\mathbb{Z}^+$ as the set of positive integers. Let $\sum$ denote the summation over $\mathbb{R}$ and $\bigoplus$ denote the summation over $\mathbb{F}_{p}$. A function $g:\mathbb{F}_p\times\cdots\times\mathbb{F}_p\to\mathbb{F}_p$ is called $\mathbb{F}_p$-linear if $g$ is a linear function with respect to $\mathbb{F}_p$. 
Random variables and their realizations are represented by uppercase letters (e.g., $S$) and lowercase letters (e.g., $s$), respectively. We use calligraphic symbols (e.g., $\mathcal{S}$) to denote sets.

Throughout the paper, all logarithms are to base two. Let $h_2(p):=-p\log(p)-(1-p)\log(1-p)$ for $p\in[0,1]$ and $0\log(0) :=0$ by convention. We denote $[1:M] := \{1,2,\cdots,M\}$, $\mathcal{A}\backslash\mathcal{B} :=\{x\in\mathcal{A}|x\notin\mathcal{B}\}$, and $\log^+(x) := \max\{\log(x),0\}$. Let $|\cdot|$ denote the cardinality of a set and $\mathbf{1}_{(\cdot)}$ denote the indicator function of an event. Given any sequence or vector $(a_1,\cdots,a_M)$ and $\mathcal{J} \subseteq [1:M]$, we denote $a_{\mathcal{J}} = (a_i:i\in\mathcal{J})$. Given any function $f$ and vectors $\mathbf{s}_i=(s_i[1],\cdots,s_i[k])$, $i\in[1:M]$, we denote $f(\mathbf{s}_1,\cdots,\mathbf{s}_M) =(f(s_1[1],\cdots,s_M[1]) , \cdots , f(s_1[k],\cdots,s_M[k]))$. Given two functions $f(x)$ and $g(x)$, we say that $f(x)=\Theta(g(x))$ if there exists $k_1, k_2>0$ and $x_0$ such that for all $x>x_0$, $k_1g(x) \le f(x) \le k_2g(x)$.

\section{Problem Statement} \label{sec:prob_state}
We consider distributed computation of a class of functions over collocated networks. The problem consists of the following basic elements: 
\begin{itemize}
\item a network consisting of $M$ sensors labeled from $1$ to $M$ and a single fusion center labeled $0$, 
\item a set of $M$ sources, each of which is observed by a unique sensor, 
\item a function $f$, which is to be computed by the fusion center, 
\item a joint source--channel code for each sensor node,
\item a decoder for the fusion center.
\end{itemize}
We now provide the mathematical definitions for each element.

\begin{definition}[Sources]
Each sensor node (indexed by $m \in[1:M]$) observes a length-$k$ vector of source symbols $\mathbf{s}_m=(s_m[1], \cdots, s_m[k])\in[0:q-1]^{k}$ which are independently drawn from the probability mass function (PMF) $p_{S_m}$, where $q\ge 2$. We assume independent source distributions, i.e.,  $p_{S_1,S_2,\cdots,S_M}=\prod_{m=1}^M p_{S_m}$.
\end{definition}

In this paper, we are interested in the following two network models. We assume a full-duplex scenario in which each node can transmit and receive simultaneously. 

\begin{figure}[t!]
\begin{center}
\includegraphics[scale=0.6]{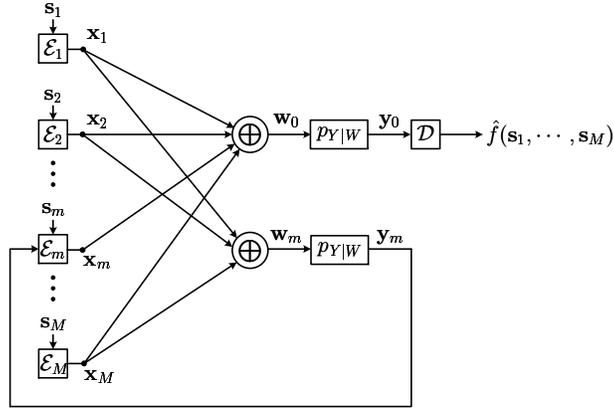}
\end{center}
\vspace{-0.15in}
\caption{Function computation in the collocated linear finite field network. Each node observes a noisy modulo-$p$ sum of transmit signals from all other nodes. To avoid clustering of lines, the figure only shows the situation of the fusion center and sensor node $m$.} 
\label{fig:LFF_network}
\vspace{-0.1in}
\end{figure}

\begin{definition}[Collocated Linear Finite Field Network]
The channel is discrete memoryless and governed by a conditional PMF
\begin{align}\label{eq:def_LFF}
p_{Y_{[0:M]}|X_{[1:M]}}(y_{[0:M]}|x_{[1:M]}) = \prod_{i=0}^Mp_{Y|W}\left(y_i|w_i\right),
\end{align}
where
\begin{align}
W_i = \bigoplus_{m\in[1:M]\backslash\{i\}} X_m, 
\end{align}
with $X_{[1:M]}\in \mathbb{F}_p^M$ and $Y_{[0:M]}\in\mathbb{F}_p^{M+1}$. Note that we assume that each multiple-access component follows the same channel law $p_{Y|W}$. An illustration of the collocated linear finite field network is given in Figure \ref{fig:LFF_network}. For convenience, let $p_{W^*}$ be one distribution achieving $\max_{p_W}I(W;Y)$.
\end{definition}

\begin{figure}[t!]
\begin{center}
\includegraphics[scale=0.7]{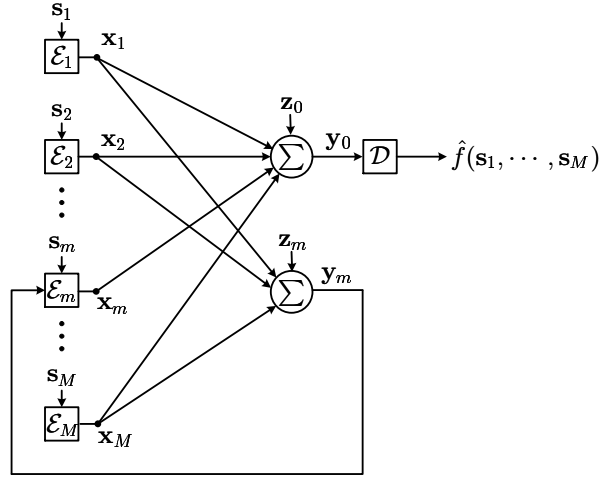}
\end{center}
\vspace{-0.15in}
\caption{Function computation in the collocated Gaussian network. Each node observes a noisy linear combination of transmit signals from all other nodes. To avoid clustering of lines, the figure only shows the situation of the fusion center and sensor node $m$.}
\label{fig:G_network}
\vspace{-0.1in}
\end{figure}

\begin{definition}[Collocated Gaussian Network]
Each node $i\in[0:M]$ observes a noisy linear combination of the transmit signals through the memoryless channel
\begin{align}\label{eq:Gchannel}
y_i = \sum_{m\in[1:M]\backslash \{i\}} x_m + z_i,
\end{align}
where $x_{[1:M]}\in\mathbb{R}^M$ and the elements of $z_{[0:M]}$ are independently drawn from $\mathcal{N}(0,1)$. An illustration of the collocated Gaussian network is given in Figure \ref{fig:G_network}.
\end{definition}

The fusion center wishes to compute a symbol-by-symbol function of the $M$ sources. In this paper, we consider the class of {\it type-threshold functions}, which is a subclass of symmetric functions. 

\begin{definition}[Symmetric Function]
Let $\Lambda$ be a finite alphabet. A function $f: [0:q-1]^M\to \Lambda$ is called symmetric if 
\begin{align}
f(s_{\sigma(1)},s_{\sigma(2)},\cdots,s_{\sigma(M)}) = f(s_1,s_2,\cdots,s_M), 
\end{align}
for every permutation $\sigma$ on $[1:M]$. 
\end{definition}

\begin{definition}[Type, Frequency Histogram]
The type (or frequency histogram) of a sequence $s_{[1:M]}\in[0:q-1]^M$ is a length-$q$ vector $b_{[0:q-1]}$ with 
\begin{align}\label{eq:def_freq}
b_{\ell} := \sum_{m=1}^M \mathbf{1}_{\{s_m=\ell\}}.
\end{align}
The $b_\ell$ is termed {\it frequency} of $\ell$.
\end{definition}

\begin{definition}[Type-Threshold Function] \label{def:TT}
Let $\Lambda$ be a finite alphabet. Let $\{f_M\}_{M\in\mathbb{Z}^+}$ be a sequence of symmetric functions, where $f_M: [0:q-1]^M\to \Lambda$ satisfies 
\begin{align}
f_M(s_1,s_2,\cdots,s_m,0,\cdots,0)=f_m(s_1,s_2,\cdots,s_m), 
\end{align}
for all $m\in[1:M]$. We say that the sequence $\{f_M\}_{M\in\mathbb{Z}^+}$ belongs to the class of type-threshold functions if there exists a non-negative integer vector $\theta_{[0:q-1]}$ and a function $g:[0:\theta_0]\times[0:\theta_1]\times\cdots\times[0:\theta_{q-1}]\to\Lambda$ such that for all $M\in\mathbb{Z}^+$, 
\begin{align}
f_M(s_1,s_2,\cdots,s_M) = g(\overline{b}_0,\cdots,\overline{b}_{q-1}),
\end{align}
where $\overline{b}_\ell := \min\left\{\theta_\ell,b_\ell\right\}$ for all $\ell \in[0:q-1]$. 
The vector $\theta_{[0:q-1]}$ is called {\it threshold vector} and $\overline{b}_\ell$ is called {\it clipped frequency} of $\ell$. In the sequel, we will simply write $f$ and the number of arguments $M$ will be clear from context.
\end{definition}

Some common instances of type-threshold functions are 
\begin{enumerate}
\item the maximum, with a threshold vector $(0, 1 , \cdots , 1)$;
\item the number of distinct elements, with a threshold vector $(1, 1 , \cdots , 1)$;
\item the average of the $\ell$ largest values, with a threshold vector $(0 , \ell , \cdots, \ell)$;
\item the frequency indicator $\mathbf{1}_{\{\exists m\in[1:M] \text{ s.t. } s_m=\ell\}}$, with a threshold vector $(0, \cdots 0, 1, 0, \cdots, 0)$ (the $1$ is on the $\ell$-th position);
\item the list of heavy hitters $\{\ell\in[0:q-1] | b_\ell\ge T\}$, with a threshold vector $(T, T , \cdots , T)$.
\end{enumerate}
Note that while the average of the $\ell$ largest values is a type-threshold function, the average $\frac{1}{M}\sum_{m=1}^Ms_m$ is not.

In the following, we give the definitions of code, rate, and capacity for the problem of function computation. 

\begin{definition}[Computation Code]
A $(k,n)$ block code for function computation is defined as
\begin{itemize}
\item (Sensor Node Encoding)  At time $i\in[1:n]$, sensor node $m\in[1:M]$ broadcasts $x_{m}[i]=\mathcal{E}_{m}^{(i)}\left(\mathbf{s}_m,\mathbf{y}_m^{i-1}\right)$.
\item (Fusion Center Decoding) The fusion center estimates $\hat{f}\left(\mathbf{s}_1,\cdots,\mathbf{s}_M\right)=\mathcal{D}\left(\mathbf{y}_0\right)$.
\end{itemize}
Here $\mathbf{y}_m^{i-1}$ denotes the length-$(i-1)$ vector containing the first $i-1$ elements of $\mathbf{y}_m$.
If the computation code is for collocated Gaussian networks, it is additionally required that each transmit signal satisfies the average power constraint $P$, i.e., $\frac{1}{n}\|\mathbf{x}_{m}\|^2\le P$. 
\end{definition}

\begin{definition}[Computation Rate]
We say that a {\it computation rate} $R:=\frac{k}{n}$ for function $f$ is achievable if there exists a sequence of $(k,n)$ computation codes such that the probability of error 
\begin{align}
\P_e^{(n)}:=\P\left(\hat{f}\left(\mathbf{s}_1,\cdots,\mathbf{s}_M\right)\neq f\left(\mathbf{s}_1,\cdots,\mathbf{s}_M\right)\right)
\end{align}
converges to zero as $n$ tends to infinity. Note that the computation rate is the number of reliably computed functions per channel use. 
\end{definition}

\begin{definition}[Computation Capacity]
The computation capacity $C$ is the supremum over all achievable computation rates. 
\end{definition}

\section{Round-Robin Broadcast with Interactive Source Coding} \label{sec:IRR}
The interactive round-robin approach follows from the framework of interactive source coding \cite{Ma:12}. The whole communication consists of $N$ rounds, where $N\ge M$. Fix a mapping $\kappa:[1:N]\to[1:M]$. In each round (indexed by $\ell\in[1:N]$), only sensor node $\kappa(\ell)$ is activated. The activated sensor $\kappa(\ell)$ quantizes the length-$k$ source vector $\mathbf{s}_{\kappa(\ell)}$ into a length-$n$ vector $\mathbf{v}_{\ell}$ with side information $\mathbf{v}_{[1:\ell-1]}$ received in previous rounds and then broadcasts this common description $\mathbf{v}_{\ell}$ to all other nodes in the network. After $N$ rounds, the fusion center computes the desired function based on the received $N$ descriptions. 
The minimum source coding rate for function computation is characterized in \cite[Corollary 1]{Ma:12}, which is stated in the following theorem.\footnote{For a formal definition, we refer the readers to Definition $1$ and $2$ in \cite{Ma:12}.}
\begin{theorem}[Ma, Ishwar, and Gupta] \label{thm:IRR}
For all $N \ge M$, the minimum source coding rate for computation of the function $f$ is 
\begin{align}\label{eq:sumrate}
\min_{p_{V_{[1:N]}|S_{[1:M]}}} I(S_{[1:M]};V_{[1:N]}),
\end{align}
where $p_{V_{[1:N]}|S_{[1:M]}}$ satisfies 
\begin{enumerate}
\item $H(f(S_{[1:M]})|V_{[1:N]})=0$, 
\item $V_\ell \leftrightarrow(V_{[1:\ell-1]},S_{\kappa(\ell)})\leftrightarrow S_{[1:M]\backslash\{\kappa(\ell)\}}$ forms a Markov chain, where $\kappa(\ell) \in[1:M]$. 
\end{enumerate}
\end{theorem}
\begin{remark}
The cardinalities of the alphabets of the descriptions $V_{[1:N]}$ can be upper bounded by functions of $q$ and $N$ without changing the minimum source coding rate for computation of the function $f$. Although we focus on type-threshold functions in this paper, the interactive round-robin approach is applicable to any function of independent discrete sources.
\end{remark}
Note that 
\begin{align}\label{eq:chain}
I(S_{[1:M]};V_{[1:N]})=\sum_{\ell=1}^N  I(S_{\kappa(\ell)};V_\ell|V_{[1:\ell-1]})
\end{align}
and intuitively we can interpret $I(S_{\kappa(\ell)};V_\ell|V_{[1:\ell-1]})$ as the source coding rate of $V_\ell$. For convenience, let $p_{V^*_{[1:N]}|S_{[1:M]}}$ be one distribution achieving \eqref{eq:sumrate} and let $V_{[1:N]}^*$ be the corresponding induced random variables.

Based on the framework of interactive source coding, we extend the achievability of the interactive round-robin approach to collocated linear finite field networks and collocated Gaussian networks. The basic idea is: First convert the networks into bit pipes with broadcast using capacity-achieving codes for point-to-point channels and then apply the interactive source coding.

\subsection{Collocated Linear Finite Field Networks} 
\begin{proposition}
In the collocated linear finite field network, any computation rate $R$ satisfying 
\begin{align}
R < \frac{I(W^*;Y)}{I(S_{[1:M]};V^*_{[1:N]})}
\end{align}
is achievable.
\end{proposition}
\begin{IEEEproof}
Denote by $n_\ell$ the number of time slots assigned to transmit the length-$k$ vector $\mathbf{v}^*_\ell$. We first compress the $\mathbf{v}^*_\ell$ into $k I(S_{\kappa(\ell)};V^*_\ell|V^*_{[1:\ell-1]})$ bits and then apply a point-to-point capacity-achieving code for channel $p_{Y|W}$. In round $\ell\in[1:N]$, the vector $\mathbf{v}^*_\ell$ sent by node $\kappa(\ell)$ can be decoded at all nodes with vanishing probability of error as $n_\ell$ increases if 
\begin{align}
k I(S_{\kappa(\ell)};V^*_\ell|V^*_{[1:\ell-1]}) < n_\ell I(W^*;Y).
\end{align}
After receiving the vectors $\mathbf{v}^*_{[1:N]}$, the fusion center can deduce the desired function as guaranteed by Theorem \ref{thm:IRR}.
Thus, by setting $n_\ell > \frac{k I(S_{\kappa(\ell)};V^*_\ell|V^*_{[1:\ell-1]})}{I(W^*;Y)}$ for all $\ell\in[1:N]$, we can achieve any computation rate $R$ satisfying  
\begin{align}
R = \frac{k}{\sum_{\ell=1}^Nn_\ell} < \frac{I(W^*;Y)}{I(S_{[1:M]};V^*_{[1:N]})}, 
\end{align}
where we used \eqref{eq:chain}.
\end{IEEEproof}

\subsection{Collocated Gaussian Networks}
Denote by $\mathcal{J}_m=\{\ell\in[1:N]\big|\kappa(\ell) = m\}$ the rounds in which node $m$ is activated. Then, we have the following proposition. 
\begin{proposition}
In the collocated Gaussian network, any computation rate $R$ satisfying 
\begin{align}\label{eq:lower_IRR}
R < \min_{\ell\in[1:N]}\frac{\frac{\alpha_\ell}{2}\log(1+P_\ell)}{I(S_{\kappa(\ell)};V^*_\ell|V^*_{[1:\ell-1]})}
\end{align}
is achievable, where $\alpha_\ell\ge 0$ and $P_\ell\ge 0$ satisfying $\sum_{\ell\in[1:N]} \alpha_\ell =1$ and $\sum_{\ell \in\mathcal{J}_m} \alpha_\ell P_\ell \le P$ for all $m\in[1:M]$.
\end{proposition}

\begin{IEEEproof}
Denote by $n_\ell$ the number of time slots assigned to transmit the length-$k$ vector $\mathbf{v}^*_\ell$ and by $P_\ell$ the corresponding transmit power. To satisfy the average power constraint, we must have for all $m\in[1:M]$, 
\begin{align}
\sum_{\ell \in\mathcal{J}_m} n_\ell P_\ell \le nP.
\end{align}

We first compress the $\mathbf{v}^*_\ell$ into $k I(S_{\kappa(\ell)};V^*_\ell|V^*_{[1:\ell-1]})$ bits and then apply a capacity-achieving code for the point-to-point Gaussian channel. In round $\ell\in[1:N]$, the vector $\mathbf{v}^*_\ell$ can be decoded at all nodes with vanishing probability of error as $n_\ell$ increases if 
\begin{align}\label{eq:IRR_condG}
k I(S_{\kappa(\ell)};V^*_\ell|V^*_{[1:\ell-1]}) < n_\ell \frac{1}{2}\log(1+P_\ell).
\end{align}
Then, we have 
\begin{align}
R = \frac{n_\ell}{n}\frac{k}{n_\ell} < \frac{n_\ell}{n}\frac{\frac{1}{2}\log(1+P_\ell)}{I(S_{\kappa(\ell)};V^*_\ell|V^*_{[1:\ell-1]})}
\end{align}
for all $\ell\in[1:N]$. Denoting $\alpha_\ell=n_\ell/n$, any computation rate  
\begin{align}
R  <  \min_{\ell\in[1:N]}\frac{\frac{\alpha_\ell}{2}\log(1+P_\ell)}{I(S_{\kappa(\ell)};V^*_\ell|V^*_{[1:\ell-1]})}
\end{align}
is achievable, where $\sum_{\ell\in[1:N]} \alpha_\ell =1$ and $\sum_{\ell \in\mathcal{J}_m} \alpha_\ell P_\ell \le P$ for all $m\in[1:M]$. 
\end{IEEEproof}
If there is no power control, i.e., setting $P_\ell=P$ and $\alpha_\ell=\frac{I(S_{\kappa(\ell)};V^*_\ell|V^*_{[1:\ell-1]})}{I(S_{[1:M]};V^*_{[1:N]})}$ in \eqref{eq:lower_IRR}, then we have 
\begin{align}
R  <  \frac{\frac{1}{2}\log(1+P)}{I(S_{[1:M]};V^*_{[1:N]})}.
\end{align} 
On the other hand,  the following theorem shows an upper bound for the interactive round-robin approach in the collocated Gaussian network. 
\begin{theorem}\label{thm:upper_IRR}
In the collocated Gaussian network, any computation rate $R$ achieved by the interactive round-robin approach must satisfy
\begin{align}\label{eq:IRR_upper}
R \le \frac{\frac{1}{2}\log(1+MP)}{I(S_{[1:M]};V^*_{[1:N]})}.
\end{align}
\end{theorem}
\begin{IEEEproof}
We refer to Appendix A for the proof.
\end{IEEEproof}

\begin{remark}
In general, the upper bound \eqref{eq:IRR_upper} cannot be achieved by optimizing over $\alpha_{[1:N]}$ and $P_{[1:N]}$ in \eqref{eq:lower_IRR} alone and an optimization over all distributions achieving \eqref{eq:sumrate} is necessary. A sufficient condition to achieve \eqref{eq:IRR_upper} is that 
\begin{enumerate}
\item $N$ is divisible by $M$,
\item $I(S_{\kappa(\ell)};V^*_\ell|V^*_{[1:\ell-1]})=\frac{1}{N}I(S_{[1:M]};V^*_{[1:N]})$ for all $\ell\in[1:N]$.
\end{enumerate}
Then, the upper bound can be achieved by setting $\alpha_\ell=\frac{1}{N}$ and $P_\ell=MP$ for all $\ell\in[1:N]$. 
\end{remark}

\section{Descriptions of the Clipped Frequencies}\label{sec:entropy}
In Theorem \ref{thm:IRR}, the auxiliary random variables $\{V_\ell\}$ describing the desired function are generated and transmitted one by one in each round. Therefore, Theorem \ref{thm:IRR} presumes a round-robin approach and its extension to collocated broadcast--superposition networks can only explore the broadcast property but not the superposition property. In this section, we propose another set of auxiliary random variables $\{U_m^{(\ell)}\}$ serving as important building blocks of the proposed multi-round group broadcast which is elaborated in the next section. Intuitively, we want to use the superposition property to somehow {\it merge} the descriptions so that the amount of information received at the receivers is reduced but still enough to deduce the desired function. 

A simple first attempt is to generalize the descriptions in Theorem \ref{thm:IRR}: Fix $N\in\mathbb{Z}^+$. Consider the descriptions $\{V_{m}^{(\ell)}\}_{m\in[1:M],\ell\in[1:N]}$ satisfying   
\begin{enumerate}
\item $H(f(S_{[1:M]})|U_{[1:N]})=0$, 
\item $V_m^{(\ell)} \leftrightarrow(U_{[1:\ell-1]},S_m)\leftrightarrow S_{[1:M]\backslash\{m\}}$ forms a Markov chain, 
\end{enumerate}
where
\begin{align}
U_\ell = \sum_{m=1}^M V_m^{(\ell)}, 
\end{align}
in which the superposition is embedded. Note that if we set $V_m^{(\ell)}=0$ for all $m\in[1:M]\backslash\{\kappa(\ell)\}$, then we recover the descriptions in Theorem \ref{thm:IRR}. These descriptions are very general but seem hard to analyze. Instead, we next propose a more constrained set of auxiliary random variables (descriptions). Not only can these descriptions be analyzed, they also have a natural operational meaning.

Rather than generating descriptions directly for the desired type-threshold function, we construct descriptions for the clipped frequencies. The reasons are twofold. First, the clipped frequencies contain all the information needed to deduce the desired type-threshold function. Second, as can be seen in \eqref{eq:def_freq}, the clipped frequencies are sums of indicators with a clipping. Thus, the indicators can serve as descriptions and the addition can play the role of merge, which is naturally matched with the superposition property of broadcast--superposition networks. In order to reduce the entropy of the descriptions, it might be unwise to attain the whole frequency and then do the clipping. Instead, we consider a recursive approach: Update only a partial sum of indicators and perform the clipping on a regular basis. Now come the details. 

First, for each $\ell\in[0:q-1]$, we attribute a partition of $[1:M]$: $\mathcal{A}_1^{(\ell)}, \cdots, \mathcal{A}_{J_\ell}^{(\ell)}$, which satisfies that 
1) $\mathcal{A}_j^{(\ell)} \neq \emptyset$ for all $j$, 2) $\bigcup_j\mathcal{A}_j^{(\ell)}=[1:M]$, 3) $\mathcal{A}_i^{(\ell)}\bigcap \mathcal{A}_j^{(\ell)}=\emptyset$ for all $i\ne j$. The sensors with index in the same set $\mathcal{A}_m^{(\ell)}$ form a group. 
Note that the formation of the groups can be different for each $\ell$. Each group is responsible for a partial sum of indicators.

Denote by $U_{1}^{(\ell)},U_{2}^{(\ell)},\cdots$ the descriptions of the clipped frequency $\overline{B}_\ell$, $\ell\in[0:q-1]$. Then, the descriptions of the clipped frequency $\overline{B}_\ell$ is defined by the following recursion
\begin{align}
\label{eq:aux}
 U_m^{(\ell)} = U_{m-1}^{(\ell)} + \sum_{i\in\mathcal{A}_{m}^{(\ell)}} \mathbf{1}_{\{U_{m-1}^{(\ell)}<\theta_\ell\}\bigcap\{S_i=\ell\}},
\end{align}
for all $m\in[1:J_\ell]$, where $U_0^{(\ell)} = 0$. As can be seen, $\sum_{i\in\mathcal{A}_{m}^{(\ell)}} \mathbf{1}_{\{S_i=\ell\}}$ is the partial sum of indicators just mentioned and the event $\{U_{m-1}^{(\ell)}<\theta_\ell\}$ plays the role of clipping. Note that $U_{[1:J_\ell]}^{(\ell)}$ are random variables induced by the sources $S_{[1:M]}$. 
It is clear that the clipped frequency $\overline{B}_\ell$ is equal to $\min\{U_{J_\ell}^{(\ell)},\theta_\ell\}$ and thus the fusion center can deduce the desired function once it learns all descriptions $\left(U_{[1:J_0]}^{(0)},U_{[1:J_1]}^{(1)},\cdots,U_{[1:J_{q-1}]}^{(q-1)}\right)$. 

\subsection{Entropy of Descriptions}\label{subsec:bin_max}
As will be clear in Section \ref{sec:MRGB}, the entropy of the descriptions $\left(U_{[1:J_0]}^{(0)},U_{[1:J_1]}^{(1)},\cdots,U_{[1:J_{q-1}]}^{(q-1)}\right)$ determines the achievable computation rate of the proposed scheme and we want this entropy to be as small as possible. In particular, we are interested in how this entropy scales as the number of sensors increases since it directly affects the scaling law of the achievable computation rate. Since the entropy of the descriptions is governed by the chosen partitions, the goal is to characterize a pattern of partitions which results in a bounded entropy as the number of sensors increases. For this, we will consider different distribution ensembles, which are families of probability distributions $\{p_{S_1}p_{S_2}\cdots p_{S_M}\}_{M\in\mathbb{Z}^+}$. 
Let us start with an example.

\begin{figure}[t!]
\centering
\includegraphics[scale=0.6, trim=0.5cm 0 0 0.3cm, clip]{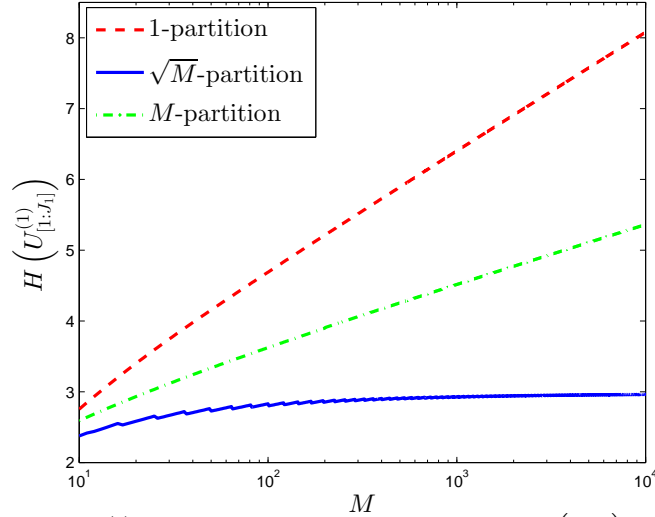}
\vspace{-0.2in}
\caption{The entropy of the descriptions $U_{[1:J_1]}^{(1)}$ (Expression \eqref{eq:HU1}) for the i.i.d. Bernoulli$\left(\frac{1}{\sqrt{M}}\right)$ ensemble under various partitions.}
\label{figs:TTCF_example}
\vspace{-0.1in}
\end{figure}

\begin{example}[Binary Maximum]
Assume that $S_m\in\{0,1\}$ for all $m\in[1:M]$. The binary maximum is defined as $S_{\max}:=\max S_{[1:M]}$. Intuitively, if we know that one sensor observes a value of one, then the function value can already be determined even though the observations at other sensors are unknown. Note that $(\theta_0,\theta_1) = (0,1)$ is a valid threshold vector of the binary maximum and thus $U_m^{(0)}=0$ for all $m$. 

We consider independent and identically distributed (i.i.d.) ensembles with Bernoulli($\beta$), where $0<\beta<1$ and $\beta$ might depend on $M$. For convenience, we use the term $a$-partition, where $a\in[1:M]$, to refer to any partition satisfying $|\mathcal{A}_j|= a$ for all $j\in[1:J-1]$ and $|\mathcal{A}_J|=M-(J-1)a$, where $J=\lfloor M/a\rfloor$. The entropy of the descriptions $U_{[1:J_1]}^{(1)}$ under $a$-partitions can be evaluated as 
\begin{align}
H\left(U_{[1:J_1]}^{(1)}\right) &\overset{(a)}{=}\sum_{m=1}^{J_1} H\left(U_m^{(1)}\big|U_{m-1}^{(1)}\right)\\
&\overset{(b)}{=}\sum_{m=1}^{J_1} \P(U_{m-1}^{(1)}=0) H\left(U_m^{(1)}\big|U_{m-1}^{(1)}=0\right)\\
\label{eq:HU1}
&\overset{(c)}{=} \frac{1-(1-\beta)^{(J_1-1)a}}{1-(1-\beta)^a}H(Q_a) + (1-\beta)^{J_1-1}H(Q_{M-(J_1-1)a})
\end{align}
where $Q_m\sim\text{Binomial}(m,\beta)$, $(a)$ follows from the independence of $S_{[1:M]}$, $(b)$ follows since $U_m^{(1)}$ conditioned on $\{U_{m-1}^{(1)}\ge 1\}$ is deterministic, and $(c)$ follows since $\P\left(U_{m-1}^{(1)}=0\right)=(1-\beta)^{(m-1)a}$,  $H\left(U_m^{(1)}\big|U_{m-1}^{(1)}=0\right)=H\left(\sum_{i\in\mathcal{A}_m^{(\ell)}}\mathbf{1}_{\{S_i=1\}}\right)$, and $\sum_{i\in\mathcal{A}_m^{(\ell)}}\mathbf{1}_{\{S_i=1\}}\sim\text{Binomial}(|\mathcal{A}_m^{(\ell)}|,\beta)$. 

Now we discuss about the following three cases. 

\noindent 1) i.i.d. Bernoulli$\left(c\right)$, where $c\in(0,1)$ is a constant independent of $M$

\noindent If we fix $a=1$, then \eqref{eq:HU1} becomes 
\begin{align}\label{eq:const}
H\left(U_{[1:J_1]}^{(1)}\right) = \frac{1}{c}(1-(1-c)^{(M-1)}) h_2(c) \underset{M\to\infty}{\longrightarrow} \frac{h_2(c)}{c}.
\end{align}
For this ensemble, the simple one-at-a-time approach gives a bounded entropy of descriptions as $M$ increases. By contrast, if we substitute $a=M$ into \eqref{eq:HU1}, then $H\left(U_{[1:J_1]}^{(1)}\right) = H(Q_M) = \Theta(\log M)$. Thus, the $M$-partition fails to achieve a bounded entropy of descriptions.
 
\noindent 2) i.i.d. Bernoulli$\left(\frac{1}{M}\right)$

\noindent If we fix $a=M$, then \eqref{eq:HU1} becomes 
\begin{align}
H\left(U_{[1:J_1]}^{(1)}\right) = H(Q_M) \overset{(a)}{\le} \frac{1}{2}\log\left(2\pi e \left(1+\frac{1}{12}\right)\right) \approx 2.1, 
\end{align}
where $(a)$ follows from \cite[Theorems 7 and 8]{Harremoes:01} and \cite[equation (1)]{Adell:10}. Thus, for this ensemble, the $M$-partition achieves a bounded entropy of descriptions as $M$ increases. By contrast, if the $1$ partition is applied, then 
\begin{align}
H\left(U_{[1:J_1]}^{(1)}\right) = M\left(1-\left(1-\frac{1}{M}\right)^{(M-1)}\right) h_2\left(\frac{1}{M}\right) \ge \frac{1}{2}\log M.
\end{align}
Thus, the $1$-partition fails to achieve a bounded entropy of descriptions. 

\noindent 3) i.i.d. Bernoulli$\left(\frac{1}{\sqrt{M}}\right)$

\noindent Figure~\ref{figs:TTCF_example} plots $H\left(U_{[1:J_1]}^{(1)}\right)$ for the $1$-partition, the $\sqrt{M}$-partition, and the $M$-partition. As can be seen, as $M$ increases, only the $\sqrt{M}$-partition achieves a bounded entropy of descriptions, which will be proved in Lemma \ref{lma:suff_info}.
\end{example}

\vspace{0.5cm}

As shown in the above example, different distribution ensembles require different partitions to achieve a bounded entropy of descriptions. The following lemma shows the existence of partitions that guarantee a bounded entropy of descriptions for any type-threshold function when the sources are independent. 
\begin{lemma} \label{lma:suff_info}
Fix a threshold vector $\theta_{[0:q-1]}$ and a joint PMF $p_{S_1}\cdots p_{S_M}$. For each $\ell\in[0:q-1]$, there exists a partition such that 
\begin{align}
H\left(U_{[1:J_\ell]}^{(\ell)}\right) < \frac{5}{2}\log(1+\theta_\ell) + 12.
\end{align}
\end{lemma}
\begin{IEEEproof}
We refer to Appendix B for the proof.
\end{IEEEproof}
Using Lemma \ref{lma:suff_info}, we can upper bound the entropy of descriptions $\left(U_{[1:J_0]}^{(0)},U_{[1:J_1]}^{(1)},\cdots,U_{[1:J_{q-1}]}^{(q-1)}\right)$ achieved by the optimum partitions as 
\begin{align}\label{eq:bounded_dsp}
H\left(U_{[1:J_0]}^{(0)},U_{[1:J_1]}^{(1)},\cdots,U_{[1:J_{q-1}]}^{(q-1)}\right) \le 12q + \frac{5}{2}\sum_{\ell=0}^{q-1} \log(1+\theta_\ell), 
\end{align}
which is independent of the number of sensors.

\subsection{Tailoring to the Maximum Function}
The descriptions introduced in \eqref{eq:aux} are a general framework for every type-threshold function. However, for many functions, it is unnecessary to acquire specific values of all clipped frequencies so as to deduce the function value. The simplest example is the frequency indicators for which we only care about one single frequency. Yet another example is the maximum function. If we directly use \eqref{eq:aux}, then we need to convey $(q-1)$ clipped frequencies and the entropy of their descriptions is upper bounded by $\Theta(q)$ as shown in \eqref{eq:bounded_dsp}. However, once all nodes learn that $\overline{b}_\ell=1$, then the values of $\overline{b}_{[0:\ell-1]}$ are irrelevant since the maximum must be larger than or equal to $\ell$. 

In this subsection, we consider an adaptation of the descriptions for the maximum function based on the {\it binary search} algorithm.
Fix $\lceil \log q\rceil$ partitions: $\mathcal{A}_{[1:J_\ell]}^{(\ell)}$, $\ell\in[1:\lceil \log q\rceil]$. For each $\ell$, define the recursion 
\begin{align}
 \tilde{U}_m^{(\ell)} = \tilde{U}_{m-1}^{(\ell)} + {\sum_{i\in\mathcal{A}_{m}^{(\ell)}}} \mathbf{1}_{\{\tilde{U}_{m-1}^{(\ell)}=0\}\bigcap\{S_i\ge D_\ell\}}, 
\end{align}
where 
\begin{align}
D_\ell = \left\lceil\frac{q}{2^{\ell}}\left(1+\sum_{j=1}^{\ell-1}\mathbf{1}_{\{\tilde{U}_{J_j}^{(j)}>0\}}2^{\ell-j}\right)\right\rceil
\end{align}
is the midpoint in the $\ell$-th stage of the binary search. For example, $D_1=\lceil\frac{q}{2}\rceil$, $D_2\in\{\lceil\frac{q}{4}\rceil,\lceil\frac{3q}{4}\rceil\}$, $D_3\in\{\lceil\frac{q}{8}\rceil,\lceil\frac{3q}{8}\rceil,\lceil\frac{5q}{8}\rceil,\lceil\frac{7q}{8}\rceil\}$, and so on.
Note that $\min \{D_{\lceil \log q\rceil},q\} = \max S_{[1:M]}$. Therefore, the fusion center can deduce the maximum once it learns the sequence $\left(\tilde{U}_{[1:J_1]}^{(1)},\tilde{U}_{[1:J_2]}^{(2)},\cdots,\tilde{U}_{[1:J_{\lceil \log q\rceil}]}^{(\lceil \log q\rceil)}\right)$. Since the proof of Lemma \ref{lma:suff_info} still follows after replacing $\P(S_i=\ell)$ by $\P(S_i\ge D_\ell)$ and substituting  $\theta_\ell=1$, the entropy $H(\tilde{U}_{[1:J_\ell]}^{(\ell)})$ can be upper bounded by a constant. Thus, the entropy of the descriptions for the maximum function is reduced from $\Theta(q)$ to $\Theta(\log q)$. 

\section{Multi-Round Group Broadcast} \label{sec:MRGB}
In this section, we elaborate our developed coding scheme {\it multi-round group broadcast}. In brief, the multi-round group broadcast conveys the descriptions of clipped frequencies introduced in Section \ref{sec:entropy} over the collocated networks with broadcast and superposition properties. Before going into the details, we first give a high level summary. To explain the main idea, it suffices to consider one of the clipped frequencies $\ell$. Let the threshold $\theta_\ell$ and the partition $\mathcal{A}_{[1:J_\ell]}^{(\ell)}$ be fixed. The operations given below are performed symbol-wise.

Before transmission, each node sets up a counter with initial value zero. There are totally $J_\ell$ rounds. In round $m\in[1:J_\ell]$, all nodes in $\mathcal{A}_m^{(\ell)}$ are activated and broadcast the indicator ``$\ell$ is observed''. The transmitted indicators are superimposed by the channel. Then, every node decodes the arithmetic sum of the indicators and increment the counter by the corresponding value. If the value of every counter reaches or exceeds the threshold $\theta_\ell$, then every node learns the clipped frequency $\overline{b}_\ell = \theta_\ell$ and we can jump directly to the next frequency; otherwise, we move on to the next round.

\subsection{Computation in Collocated Linear Finite Field Networks}\label{sec:Fp}
In this subsection, we formally describe the proposed multi-round group broadcast in collocated linear finite field networks. Let the partitions $\{\mathcal{A}_m^{(\ell)}\}$ be fixed. 
In the $m$-th round of the transmission of $\overline{\mathbf{b}}_\ell$, where $m\in[1:J_\ell]$, the activated group $\mathcal{A}_m^{(\ell)}$ cooperatively informs all nodes of a length-$k$ vector $\mathbf{u}_m^{(\ell)}$ with entries
\begin{align} \label{eq:PI_entry}
u_m^{(\ell)}[j] = u_{m-1}^{(\ell)}[j] + \sum_{i\in\mathcal{A}_m^{(\ell)}} \mathbf{1}_{\{u_{m-1}^{(\ell)}[j]<\theta_\ell\}\bigcap\{s_i[j]=\ell\}}
\end{align}
for $j\in[1:k]$.
Since all nodes learn $\mathbf{u}_{m-1}^{(\ell)}$ in the previous round, the activated group only needs to cooperatively broadcast the arithmetic sum of indicators in \eqref{eq:PI_entry}. 

The problem of computing an arithmetic sum in a linear finite field multiple access channel (MAC) remains open in general, but if $p_{Y|W}$ is symmetric (see Definition 13 in \cite{Nazer:07}), the linear computation coding \cite{Nazer:07} achieves the computation capacity. The coding scheme can be easily extended to the case when the same side information is available at all nodes. 
\begin{theorem}[Nazer and Gastpar] \label{thm:LSW}
Let $g$ be an $\mathbb{F}_p$-linear function. Assume that all receivers observe side information $V$. Then, any computation rate $R$ satisfying   
\begin{align}
R < \frac{I(W;Y)}{H(g(S_{[1:M]})|V)}
\end{align}
is achievable in the collocated linear finite field network, where $I(W;Y)$ is evaluated using a uniform distribution on $\mathbb{F}_p$.
\end{theorem}
If we restrict that $|\mathcal{A}_m^{(\ell)}| < p$, for all $\ell, m$, then each partial sum of indicators is an $\mathbb{F}_p$-linear function and we have the following achievability for computation of type-threshold functions.

\begin{theorem} \label{thm:TT_F}
Consider computation of a type-threshold function with threshold vector $\theta_{[0:q-1]}$ in the collocated linear finite field network. For each $\ell\in[0:q-1]$, fix a partition satisfying $\displaystyle\max_m |\mathcal{A}_m^{(\ell)}|<p$. Then, any computation rate $R$ satisfying 
\begin{align}
\label{eq:F_rate}
R < \frac{I(W;Y)}{H\left(U_{[1:J_0]}^{(0)},U_{[1:J_1]}^{(1)},\cdots,U_{[1:J_{q-1}]}^{(q-1)}\right)},
\end{align}
is achievable, where the ${\{U_m^{(\ell)}\}}$ are given by \eqref{eq:aux} and $I(W;Y)$ is evaluated using a uniform distribution on $\mathbb{F}_p$.
\end{theorem}

\begin{IEEEproof} It suffices to show that each length-$k$ vector $\mathbf{u}_m^{(\ell)}$ (see \eqref{eq:PI_entry}) can be decoded reliably with high probability 
if the number of assigned time slots $n_m^{(\ell)}$ satisfies that  
\begin{align}\label{eq:F_condn}
n_m^{(\ell)} > \frac{kH\left(U_m^{(\ell)}\Big|U_{[1:J_0]}^{(0)},U_{[1:J_1]}^{(1)},\cdots,U_{[1:m-1]}^{(\ell)}\right)}{I(W;Y)}, 
\end{align}
since then summing up all required time slots and noticing that $R = \frac{k}{\sum_{\ell=0}^{q-1} \sum_{m=1}^{J_\ell}n_m^{(\ell)}}$ establishes \eqref{eq:F_rate}.

Upon transmission of $\mathbf{u}_m^{(\ell)}$, the side information $(\mathbf{u}_{[1:J_0]}^{(0)},\mathbf{u}_{[1:J_1]}^{(1)},\cdots,\mathbf{u}_{[1:m-1]}^{(\ell)}) $ is available at all nodes. We apply Theorem \ref{thm:LSW} by setting $V=(U_{[1:J_0]}^{(0)},U_{[1:J_1]}^{(1)},\cdots,U_{[1:m-1]}^{(\ell)})$ and 
\begin{align}
g(S_{[1:M]}) = \sum_{i\in\mathcal{A}_m^{(\ell)}} \mathbf{1}_{\{U_{m-1}^{(\ell)}<\theta_\ell\}\bigcap\{S_i=\ell\}}.
\end{align}
Then, \eqref{eq:F_condn} is established from the fact that $H(U_m^{(\ell)}|V)=H(g(S_{[1:M]})|V)$.
\end{IEEEproof}

Applying the upper bound \eqref{eq:bounded_dsp} to Theorem \ref{thm:TT_F}, we have the following corollary.
\begin{corollary}\label{col:const_F}
Consider the collocated linear finite field network with $M$ sensors, where $M<p$. Given any type-threshold function with threshold vector $\theta_{[0:q-1]}$, any computation rate $R$ satisfying  
\begin{align}
R < \frac{I(W;Y)}{12q + \frac{5}{2}\sum_{\ell=0}^{q-1} \log(1+\theta_\ell)},
\end{align}
can be achieved by the multi-round group broadcast, where $I(W;Y)$ is evaluated using a uniform distribution on $\mathbb{F}_p$.
\end{corollary}
Essentially, Corollary \ref{col:const_F} says that when the field order is much larger than the number of sensors, i.e., $p\gg M$, then the achievable computation rate of every type-threshold function by the multi-round group broadcast will be affected little when the number of sensors increases. 

\subsection{Computation in Collocated Gaussian Networks}\label{sec:Gaussian}
In this subsection, we formally describe the proposed multi-round group broadcast in collocated Gaussian networks. We first transform the collocated Gaussian network into a collocated linear finite field network. Specifically, we apply the compute-and-forward framework \cite{Nazer:11} to transform the Gaussian MAC in \eqref{eq:Gchannel} with $n$ channel uses into the following length-$t$ deterministic linear finite field MAC over $\mathbb{F}_p$: 
\begin{align}
\mathbf{y}'_i = \bigoplus_{m\in\mathcal{A}\backslash\{i\}}\mathbf{x}'_m, 
\end{align}
for all $i\in[0:M]$, where $\mathcal{A} \subseteq [1:M]$ and $\mathbf{x}'_m\in\mathbb{F}_p^t$.
Applying Theorem 4 in \cite{Nazer:11} to \eqref{eq:Gchannel} by setting $\mathbf{a}_i=(1,1,\cdots,1)$ for all $i\in[0:M]$, we have the following theorem.

\begin{theorem}[Nazer and Gastpar] \label{thm:CF}
Fix $\mathcal{A}\subseteq[1:M]$. Let $\mathbf{x}'_{\mathcal{A}}$ be independently and uniformly drawn from $\mathbb{F}_p^t$. In the collocated Gaussian network, if $\lim_{n\to \infty}\frac{n}{p} = 0$ and 
\begin{align}\label{eq:CF_rate} 
\frac{t\log p}{n} < \frac{1}{2} \log^+\left(\frac{1}{|\mathcal{A}|}+P\right), 
\end{align}
then for all $i\in[0:M]$, $\mathbf{y}'_i = \bigoplus_{m\in\mathcal{A}\backslash\{i\}}\mathbf{x}'_m$ can be computed reliably with vanishing probability of error as $n$ increases.
\end{theorem}

The condition $\lim_{n\to \infty}\frac{n}{p} = 0$ implies that this transformation leads to a linear finite field network with unbounded field order.\footnote{We refer readers to \cite[Appendix B]{Nazer:11} for details.} This feature matches the multi-round group broadcast since for $p$ large enough, any  arithmetic sum with finite support becomes a modulo-$p$ sum, i.e., an $\mathbb{F}_p$-linear function. Therefore, in collocated Gaussian networks, we can apply the multi-round group broadcast without any restriction on the partition of sensors $[1:M]$.

One main difference between the Gaussian model and the linear finite field model is the availability of power control in the Gaussian model. Specifically, since the power constraint is imposed as an average over long time horizons, each group can use larger transmit power during its active time period by accumulating power in its non-active time period.
\begin{theorem}\label{thm:rate_G}
Consider computation of a type-threshold function with threshold vector $\theta_{[0:q-1]}$ in the collocated Gaussian network. Fix a partition $\mathcal{A}_{[1:J_\ell]}^{(\ell)}$ for each $\ell\in[0:q-1]$. Then, for any values $\alpha_m^{(\ell)}\ge 0$ and $P_m^{(\ell)}\ge 0$ satisfying $\sum_{\ell=0}^{q-1}\sum_{m=1}^{J_\ell}\alpha_m^{(\ell)}\le1$ and $\sum_{(\ell,m) \text{ s.t. } i\in\mathcal{A}_m^{(\ell)}}\alpha_m^{(\ell)}P_m^{(\ell)}\le P$ for all $i\in[1:M]$, any computation rate $R$ satisfying 
\begin{align}\label{eq:rate_G}
R <\min_{\ell\in[0:q-1]}\min_{m\in{[1:J_\ell]}}\frac{\frac{\alpha_m^{(\ell)}}{2}\log^+\left(\frac{1}{|\mathcal{A}_m^{(\ell)}|}+ P_m^{(\ell)}\right)}{H\left(U_{m(Q_\ell)}^{(\ell)}\Big|U_{[1:J_0]}^{(0)},U_{[1:J_1]}^{(1)},\cdots,U_{[1:m(Q_\ell)-1]}^{(\ell)},Q_{[0:q-1]}\right)}
\end{align}
is achievable, where 
\begin{align}
\label{eq:U_rand}
U_{m(Q_\ell)}^{(\ell)} = U_{m(Q_\ell)-1}^{(\ell)} + \sum_{i\in\mathcal{A}_{m}^{(\ell)}}\mathbf{1}_{\{U_{m(Q_\ell)-1}^{(\ell)}<\theta_\ell\}\bigcap\{S_i=\ell\}}, 
\end{align}
$U_{0}^{(\ell)} = 0$, $m(Q_\ell) := ((m+Q_\ell-1) \mod J_\ell)+1$, and $Q_\ell \sim \text{Uniform}([0:J_\ell-1])$. 
\end{theorem}

\begin{IEEEproof} 
Draw $Q_\ell$ uniformly at random from $[1:J_\ell]$ for each $\ell\in[0:q-1]$. All nodes agree upon $Q_{[0:q-1]}$ before transmission. During the transmission of the descriptions of the clipped frequency $\overline{\mathbf{b}}_\ell$, the transmission order is $Q_\ell+1, Q_\ell+2,\cdots, J_\ell, 1, 2, \cdots, Q_\ell$. Assuming that $(\mathbf{u}_{[1:J_0]}^{(0)},\mathbf{u}_{[1:J_1]}^{(1)},\cdots,\mathbf{u}_{[1:m(Q_\ell)-1]}^{(\ell)}) $ are successfully decoded at all nodes, we consider the transmission of $\mathbf{u}_{m(Q_\ell)}^{(\ell)}$ by sensor node $m$. Denote by $n_m^{(\ell)}$ and $P_m^{(\ell)}$ the number of assigned time slots and the transmit power, respectively. Let $n$ denote the total block length. Clearly, we must have\footnote{Allowing $n$ to be strictly larger enables bursty transmission so as to ensure a positive rate whenever $P>0$.}
\begin{align} \label{eq:len_cs}
\sum_{\ell=0}^{q-1}\sum_{m=1}^{J_\ell}n_m^{(\ell)} \le n.
\end{align}
Also, to satisfy the average power constraint, we must have for all $i\in[1:M]$, 
\begin{align} \label{eq:pwr_cs}
\sum_{(\ell,m) \text{ s.t. } i\in\mathcal{A}_m^{(\ell)}}n_m^{(\ell)}P_m^{(\ell)}\le nP.
\end{align}
If $H\left(U_{m(Q_\ell)}^{(\ell)}\Big|U_{[1:J_0]}^{(0)},U_{[1:J_1]}^{(1)},\cdots,U_{[1:m(Q_\ell)-1]}^{(\ell)},Q_{[0:q-1]}\right)=0$, we simply set $n_m^{(\ell)} = 0$. In the following, we assume that $H\left(U_{m(Q_\ell)}^{(\ell)}\Big|U_{[1:J_0]}^{(0)},U_{[1:J_1]}^{(1)},\cdots,U_{[1:m(Q_\ell)-1]}^{(\ell)},Q_{[0:q-1]}\right)>0$.

Setting $p$ as the least prime number larger than $n\log n$, the condition $\lim_{n\to\infty}\frac{n}{p}=0$ is satisfied. Thus, Theorem \ref{thm:CF} says that we can transform length-$n_m^{(\ell)}$ Gaussian MACs into length-$t_m^{(\ell)}$ deterministic linear finite field MACs over $\mathbb{F}_p$ if  
\begin{align}\label{eq:Gcond}
t_m^{(\ell)}\log p < n_m^{(\ell)}\frac{1}{2}\log^+\left(\frac{1}{|\mathcal{A}_m^{(\ell)}|}+P_m^{(\ell)}\right).
\end{align}
Setting $n$ large enough, an arithmetic sum of indicators becomes $\mathbb{F}_p$-linear. Thus, Theorem \ref{thm:LSW} says that $\mathbf{u}_{m(Q_\ell)}^{(\ell)}$ can be computed reliably at all nodes using $t_m^{(\ell)}$ time slots if 
\begin{align} \label{eq:Fcond}
kH\left(U_{m(Q_\ell)}^{(\ell)}\Big|U_{[1:J_0]}^{(0)},U_{[1:J_1]}^{(1)},\cdots,U_{[1:m(Q_\ell)-1]}^{(\ell)},Q_{[0:q-1]}\right) < t_m^{(\ell)} \log p.
\end{align}
Combining \eqref{eq:Gcond} and \eqref{eq:Fcond} leads to 
\begin{align}\label{eq:Grate_n}
R =\frac{k}{n}<\frac{n_m^{(\ell)}}{n}\frac{\frac{1}{2}\log^+\left(\frac{1}{|\mathcal{A}_m^{(\ell)}|}+P_m^{(\ell)}\right)}{H\left(U_{m(Q_\ell)}^{(\ell)}\Big|U_{[1:J_0]}^{(0)},U_{[1:J_1]}^{(1)},\cdots,U_{[1:m(Q_\ell)-1]}^{(\ell)},Q_{[0:q-1]}\right)}.
\end{align}
Finally, \eqref{eq:rate_G} is established after we substitute $\alpha_m^{(\ell)}=n_m^{(\ell)}/n$ into \eqref{eq:len_cs}, \eqref{eq:pwr_cs}, and \eqref{eq:Grate_n} and minimize the right hand side of \eqref{eq:Grate_n} among all $m\in[1:J_\ell]$, $\ell\in[0:q-1]$.
\end{IEEEproof}

\begin{remark}\label{rmk:equi_A}
Expression \eqref{eq:U_rand} is equivalent to saying that 
\begin{align}
U_{m}^{(\ell)} = U_{m-1}^{(\ell)} + \sum_{i\in\mathcal{A}_{m(M-Q_\ell)}^{(\ell)}}\mathbf{1}_{\{U_{m-1}^{(\ell)}<\theta_\ell\}\bigcap\{S_i=\ell\}}.
\end{align}
\end{remark}

Next, we would like to gain insight into \eqref{eq:rate_G}. The denominator of \eqref{eq:rate_G} can be upper bounded as 
\begin{align}
& H\left(U_{m(Q_\ell)}^{(\ell)}\Big|U_{[1:J_0]}^{(0)},U_{[1:J_1]}^{(1)},\cdots,U_{[1:m(Q_\ell)-1]}^{(\ell)},Q_{[0:q-1]}\right) \\
\le  & H\left(U_{m(Q_\ell)}^{(\ell)}\Big|U_{m(Q_\ell)-1}^{(\ell)},Q_\ell\right) \\
=  & \frac{1}{J_\ell}\sum_{d=0}^{J_\ell-1} H\left(U_{m(d)}^{(\ell)}\Big|U_{m(d)-1}^{(\ell)},Q_\ell = d\right) \\
\label{eq:sum_HU}
=  & \frac{1}{J_\ell}\sum_{d=0}^{J_\ell-1} H\left(U_{m(d)}^{(\ell)}\Big|U_{m(d)-1}^{(\ell)}\right).
\end{align}
The following lemma shows the existence of partitions that guarantees \eqref{eq:sum_HU} is bounded. Note that Lemma \ref{lma:suff_info} is a special case of Lemma \ref{lma:suff_d} with $d=0$.
\begin{lemma}\label{lma:suff_d}
Fix a threshold vector $\theta_{[0:q-1]}$ and a joint PMF $p_{S_1}\cdots p_{S_M}$. For each $\ell\in[0:q-1]$, there exists a partition such that for any $d\in[0:J_\ell-1]$, 
\begin{align}\label{eq:upper_Q}
\sum_{m=0}^{J_\ell-1} H\left(U_{m(d)}^{(\ell)}\Big|U_{m(d)-1}^{(\ell)}\right) < \frac{5}{2}\log(1+\theta_\ell) + 12, 
\end{align}
where $U_{m(d)}^{(\ell)}$ is given by \eqref{eq:U_rand} with $Q_\ell=d$.
\end{lemma}
\begin{IEEEproof}
We refer to Appendix B for the proof.
\end{IEEEproof}
If we use Lemma \ref{lma:suff_d} to lower bound the rate given in Theorem \ref{thm:rate_G} (Equation \eqref{eq:rate_G}), we obtain the following corollary:
\begin{corollary} \label{col:const_G}
Consider computation of a type-threshold function with threshold vector $\theta_{[0:q-1]}$ in the collocated Gaussian network. Any computation rate $R$ satisfying  
\begin{align}
R < \max_{\beta\in(0,1]}\frac{\frac{\beta}{2}\log^+\left(\frac{1}{M}+ \frac{\min_{\ell\in[0:q-1]}J_\ell}{\beta}P \right)}{12q + \frac{5}{2}\sum_{\ell=0}^{q-1} \log(1+\theta_\ell)},
\end{align}
can be achieved by the multi-round group broadcast, where $J_{[0:q-1]}$ are determined by the partition used in the proof of Lemma \ref{lma:suff_d}.
\end{corollary}
\begin{IEEEproof}
First, combining Theorem \ref{thm:rate_G}, Expression \eqref{eq:sum_HU}, and Lemma \ref{lma:suff_d} shows that any computation rate $R$ satisfying 
\begin{align}
R < \min_{\ell\in[0:q-1]}\min_{m\in{[1:J_\ell]}}\frac{\frac{\alpha_m^{(\ell)}}{2}\log^+\left(\frac{1}{|\mathcal{A}_m^{(\ell)}|}+ P_m^{(\ell)}\right)}{\frac{1}{J_\ell} \left(\frac{5}{2}\log(1+\theta_\ell) + 12\right)}
\end{align}
is achievable. Then, setting $\alpha_m^{(\ell)}=\beta\alpha_\ell/J_\ell$, $P_m^{(\ell)}=J_\ell P/\beta$ and noticing $|\mathcal{A}_m^{(\ell)}| \le M$ gives 
\begin{align}
R < \min_{\ell\in[0:q-1]}\frac{\frac{\beta\alpha_\ell}{2}\log^+\left(\frac{1}{M}+ \frac{J_\ell}{\beta} P\right)}{\frac{5}{2}\log(1+\theta_\ell) + 12}, 
\end{align}
where $\alpha_\ell \ge 0$, $\sum_{\ell=0}^{q-1}\alpha_\ell\le1$, and $\beta\in(0,1]$.
Finally, we set 
\begin{align}
\alpha_\ell = \frac{\frac{5}{2}\log(1+\theta_\ell) + 12}{12q + \frac{5}{2}\sum_{i=0}^{q-1}\log(1+\theta_i)}, 
\end{align}
and thus we have 
\begin{align}\label{eq:prerate_G}
R < \frac{\frac{\beta}{2}\log^+\left(\frac{1}{M}+ \frac{\min_{\ell\in[0:q-1]}J_\ell}{\beta}P \right)}{12q + \frac{5}{2}\sum_{i=0}^{q-1}\log(1+\theta_i)}. 
\end{align}
The corollary is established after we maximize the right hand side of \eqref{eq:prerate_G} over $\beta\in(0,1]$.
\end{IEEEproof}
This corollary establishes two key facts. First, even if the number of sensors tends to infinity, the multi-ground group broadcast still achieves a positive rate as long as $P>0$. Second, depending on the source distribution, the achievable computation rate can even increase with the number of sensors through the gain $\min_{\ell\in[0:q-1]}J_\ell$. 

\subsection{Scaling Law for the Number of Sensors and the Transmit Power: Binary Maximum}
In this subsection, we study the interplay between the number of sensors and the transmit power in the collocated Gaussian network. We consider the binary maximum function introduced in Section \ref{subsec:bin_max}. For the interactive round-robin approach, the achievable computation rate \eqref{eq:IRR_upper} can be further bounded by applying the first bound of Theorem 3 in \cite{Ma:12} to the binary maximum: 
\begin{align} 
R \le \frac{\frac{1}{2}\log(1+MP)}{Mh_2(\alpha) -(M-1)\left(1-(1-\alpha)^M\right)h_2\left(\frac{\frac{M\alpha}{1-(1-\alpha)^M}-1}{M-1}\right)}.
\end{align}
For the multi-round group broadcast, the rate expression \eqref{eq:rate_G} can be simplified as
\begin{align}\label{eq:bin_G}
R < \min_{m\in[1:J]}\frac{\frac{1}{2}\log^+\left(\frac{1}{|\mathcal{A}_m^{(1)}|}+J P\right)}{\sum_{d=0}^{J-1} H\left(U_{m(d)}^{(1)}\Big|U_{m(d)-1}^{(1)}\right)}.
\end{align}

Again, we consider the following three distribution ensembles.

\noindent 1) i.i.d. Bernoulli$\left(c\right)$, where $c\in(0,1)$ is a constant indepenent of $M$

\noindent In this ensemble, both the interactive round-robin approach and the multi-round group broadcast with the $1$-partition achieve the scaling law of $\Theta(\log MP)$. In general, the interactive round-robin approach achieves a higher computation rate than the multi-round group broadcast restricted to the $1$-partition since the latter is a special case of the former.
 
\noindent 2) i.i.d. Bernoulli$\left(\frac{1}{M}\right)$

\noindent In this ensemble, the interactive round-robin approach achieves the scaling law of $\Theta\left(\frac{\log MP}{\log M}\right)$. By contrast, the multi-round group broadcast with the $M$-partition achieves the scaling law of $\Theta(\log P)$. 

\noindent 3) i.i.d. Bernoulli$\left(\frac{1}{\sqrt{M}}\right)$

\begin{figure}[t!]
\centering
\includegraphics[scale=0.7, trim=1.5cm 0 0 0.3cm, clip]{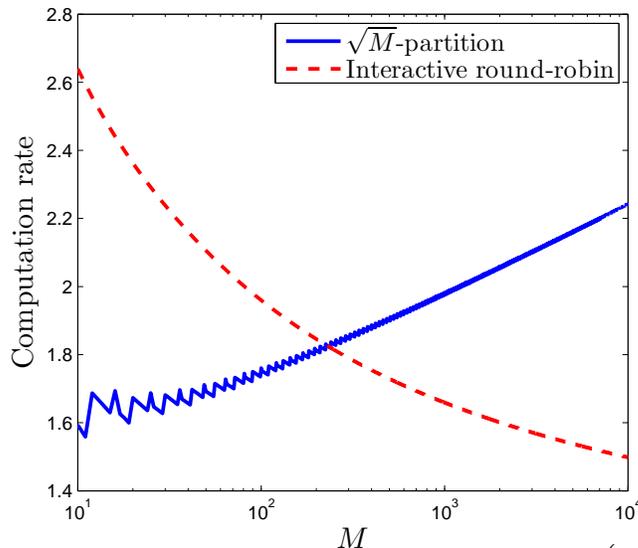}
\vspace{-0.2in}
\caption{Achievable computation rates of the binary maximum function for the i.i.d. Bernoulli$\left(\frac{1}{\sqrt{M}}\right)$ ensemble. $P=20$ dB. }
\label{figs:TTCG_example}
\vspace{-0.1in}
\end{figure}

\noindent Figure~\ref{figs:TTCG_example} plots the computation rates of the proposed multi-round group broadcast with the $\sqrt{M}$-partition and the interactive round-robin approach at $P = 20$ dB. The figure shows that as $M$ increases, the achievable computation rate of the multi-round group broadcast grows logarithmically with $M$, while the interactive round-robin approach achieves at most a constant rate. The reason that the computation rate can increase with $M$ is that, for this ensemble  the upper bound \eqref{eq:upper_Q} can be satisfied by using the $\sqrt{M}$-partition which results in an additional gain roughly $\sqrt{M}$.

\section{Upper Bound}\label{sec:cut-set}
In this section, we first provide a simple cut-set based upper bound on the computation capacity for arbitrary functions and networks. Then, we specialize this upper bound for the collocated linear finite field network and the collocated Gaussian network. In general, the derived upper bounds can not be matched by the achievabilities presented in this paper. We remark that it might be possible to tighten the upper bounds by applying the converse of interactive source coding for function computation \cite{Ma:11}.

Let $\Omega \subseteq [1:M]$ and  $\Omega^c:=[1:M]\backslash\Omega$. First, assume that a genie provides $\mathbf{s}_{\Omega^c}$ to all nodes. Given $\mathbf{s}_{\Omega^c}$ as side information, the minimum source coding rate for computation of the function $f$ should be at least $H(f(S_{[1:M]})|S_{\Omega^c})$. Second, we treat sensor nodes in $\Omega$ as a supernode-$\Omega$ to which $\mathbf{s}_\Omega$ are available. Also, we treat sensor nodes in $\Omega^c$ and the fusion center as supernode-$\{0\}\bigcup\Omega^c$. Thus, the channel from the supernode-$\Omega$ to the supernode-$\{0\}\bigcup\Omega^c$ is a point-to-point multiple-input multiple-output (MIMO) channel in which the source--channel separation theorem holds and feedback does not increase the capacity since the channel is memoryless. 
Therefore, following similar lines in the proof of the cut-set bound \cite[Theorem 15.10.1]{Cover:06}, the computation capacity of the function $f$ is upper bounded as 
\begin{align}\label{eq:cut-set bound}
C\le \max_{p_{X_{[1:M]}}} \min_{\substack{\Omega\subseteq[1:M]\\ \Omega\neq \emptyset}} \frac{I(X_{\Omega};Y_0,Y_{\Omega^c}|X_{\Omega^c})}{H(f(S_{[1:M]})|S_{\Omega^c})}, 
\end{align}
where the input distribution $p_{X_{[1:M]}}$ might subject to some constraints depending on the network model.

Specializing \eqref{eq:cut-set bound} for the two network models considered in this paper, we have the following propositions.
\begin{proposition}
In the collocated linear finite field network, the computation capacity of the function $f$ is upper bounded as 
\begin{align}
C \le \max_{p_W}\min_{\substack{\Omega\subseteq[1:M]\\ \Omega\neq \emptyset}}\frac{I(W;Y_0,Y_{\Omega^c})}{H(f(S_{[1:M]})|S_{\Omega^c})}.
\end{align}
\end{proposition}
\begin{IEEEproof}
The mutual information term in \eqref{eq:cut-set bound} can be upper bounded as 
\begin{align}
I(X_{\Omega};Y_0,Y_{\Omega^c}|X_{\Omega^c}) &= H(Y_0,Y_{\Omega^c}|X_{\Omega^c}) - H(Y_0,Y_{\Omega^c}|X_{[1:M]}) \\
& \le H(Y_0,Y_{\Omega^c}) - H(Y_0,Y_{\Omega^c}|W,X_{[1:M]}) \\
& \overset{(a)}{=} H(Y_0,Y_{\Omega^c}) - H(Y_0,Y_{\Omega^c}|W) \\
& = I(W;Y_0,Y_{\Omega^c})
\end{align}
where $W=\bigoplus_{m=1}^MX_m$ and $(a)$ follows the channel law \eqref{eq:def_LFF}. The proposition follows immediately.
\end{IEEEproof}

\begin{proposition}
In the collocated Gaussian network, the computation capacity of the function $f$ is upper bounded as 
\begin{align}\label{eq:cutset_G}
C \le \max_{\mathbf{K}}\min_{\substack{\Omega\subseteq[1:M]\\ \Omega\neq \emptyset}}\frac{\frac{1}{2}\log\left(1+(M+1-|\Omega|)\sum_{i,j}[\mathbf{K}_{X_{\Omega}|X_{\Omega^c}}]_{ij}\right)}{H(f(S_{[1:M]})|S_{\Omega^c})}, 
\end{align}
where the matrix $\mathbf{K}$ is positive semidefinite with the $(i,i)$ entry $[\mathbf{K}]_{ii}\le P$, $i\in[1:M]$ and $\mathbf{K}_{X_{\Omega}|X_{\Omega^c}}$ is the conditional covariance matrix of $X_\Omega$ given $X_{\Omega^c}$ for $X_{[1:M]}\sim\mathcal{N}(\mathbf{0},\mathbf{K})$. 
\end{proposition}

\begin{IEEEproof}
Denote by $\mathbf{K}$ the covariance matrix of $X_{[1:M]}$ with the $(i,i)$ entry $[\mathbf{K}]_{ii}\le P$, $i\in[1:M]$.
Applying Theorem 19.1 in \cite{ElGamal:11}, the mutual information term in \eqref{eq:cut-set bound} can be upper bounded as 
\begin{align}
I(X_{\Omega};Y_{\Omega^c}|X_{\Omega^c}) &\le \frac{1}{2}\log\left(\det\left(\mathbf{I} + \mathbf{G}\mathbf{K}_{X_{\Omega}|X_{\Omega^c}}\mathbf{G}^T\right)\right) \\
&= \frac{1}{2}\log\left(1+(M+1-|\Omega|)\sum_{i,j}[\mathbf{K}_{X_{\Omega}|X_{\Omega^c}}]_{ij}\right), 
\end{align}
where $\mathbf{I}$ is the $(M+1-|\Omega|)\times(M+1-|\Omega|)$ identity matrix, $\mathbf{G}$ is the $(M+1-|\Omega|)\times |\Omega|$ all-one matrix, and $\mathbf{K}_{X_{\Omega}|X_{\Omega^c}}$ is the conditional covariance matrix of $X_\Omega$ given $X_{\Omega^c}$. The equality holds if $X_{[1:M]}\sim\mathcal{N}(\mathbf{0},\mathbf{K})$.
Then, the proposition follows immediately.
\end{IEEEproof}

\begin{example}
Consider the computation of the binary maximum in the collocated Gaussian network. Assume that the distribution ensemble is i.i.d. Bernoulli$\left(\frac{1}{M}\right)$. Recall that the interactive round-robin approach achieves the scaling law of $\Theta\left(\frac{\log MP}{\log M}\right)$, whereas the multi-round group broadcast with the $M$-partition achieves the scaling law of $\Theta(\log P)$. If we just consider the cut $\Omega=[1:M]$, then the cut-set bound \eqref{eq:cutset_G} can be simplified as 
\begin{align}
C\le \frac{\frac{1}{2}\log(1+M^2P)}{h_2\left((1-1/M)^M\right)}.
\end{align}
Thus, in this ensemble the scaling of the cut-set bound \eqref{eq:cutset_G} is at most $\Theta(\log MP)$.
\end{example}

\section{Concluding Remarks} \label{sec:conclude}
In this paper, we have developed a coding scheme for computation of type-threshold functions over networks with the broadcast and superposition properties. The proposed coding scheme essentially decomposes a type-threshold function into several linear functions, which can be reliably computed over multiple-access components using computation codes. We showed that a careful scheduling of concurrent transmission is needed so as to have a bounded entropy of the clipped frequencies as the number of sensors increases. In many cases, the proposed multi-round group broadcast outperforms a combination of interactive source coding and point-to-point channel codes. 

The problem of type-threshold function computation over wireless networks remains unsolved. There are several aspects that can be pushed forward based on this work. First, to improve the achievability, one can consider other formulations of description which also explores the broadcast and superposition property of wireless networks. For the converse, we believe that a statement similar to \cite[Lemma 3]{Ma:12} can be established which says that in order to deduce the desired function, the fusion center must learn more than the function itself. Second, the multi-round group broadcast works also for correlated sources and it is desirable to know whether a constant computation rate independent of the number of sensors is still achievable. Finally, one can extend the multi-round group broadcast to general broadcast--superposition networks with multiple access components governed by various channel laws and requiring multi-hop communications. 

\section*{Appendix A \\ An Upper Bound for the Interactive Round Robin}

Before proving Theorem \ref{thm:upper_IRR}, we need the following two lemmas.

\begin{lemma}\label{lma:x}
Consider the optimization problem
\begin{align}
\text{maximize    } \hspace{0.5cm} & \min_{i\in[1:N]}\frac{\theta_i}{x_i} \\
\text{subject to    } \hspace{0.5cm} & \sum_{i=1}^N x_i = \gamma, \\
			      \hspace{0.5cm}  & x_i>0, \quad i\in[1:N],
\end{align}
where $\{\theta_i\}$ and $\gamma$ are positive constants.  An optimum solution is $x_i = \frac{\gamma\theta_i}{\sum_{j=1}^{N}\theta_j}$ for all $i\in[1:N]$ and the attained maximum is $\frac{1}{\gamma}\sum_{j=1}^{N}\theta_j$.
\end{lemma}
\begin{IEEEproof}
First, by introducing an auxiliary variable $r$, the optimization problem can be reformulated as the following:
\begin{align}
\text{maximize    }  \hspace{0.5cm} & r \\ 
\label{eq:con1} \text{subject to    }  \hspace{0.5cm} & r \le \frac{\theta_i}{x_i}, \quad i\in[1:N], \\
\label{eq:con2}				\hspace{0.5cm} & \sum_{i=1}^N x_i = \gamma, \\
				\hspace{0.5cm} & x_i>0, \quad i\in[1:N]. \nonumber 
\end{align}
It is easy to see that the optimum $r^*>0$. Notice that \eqref{eq:con1} and \eqref{eq:con2} imply that 
\begin{align}
\gamma=\sum_{i=1}^N x_i \le \frac{1}{r^*}\sum_{i=1}^N\theta_i,
\end{align}
and thus $r^*\le \frac{1}{\gamma}\sum_{i=1}^N\theta_i$. It can be easily checked that setting $x_i = \frac{\gamma\theta_i}{\sum_{j=1}^{N}\theta_j}$ for all $i\in[1:N]$ attains the upper bound. 
\end{IEEEproof}

\begin{lemma}\label{lma:y}
Let $g(\cdot)$ be a concave function defined on a real interval. Consider the optimization problem
\begin{align}
\text{maximize    } \hspace{0.5cm} & \sum_{i=1}^{N}g(y_i) \\
\text{subject to    } \hspace{0.5cm} & \sum_{i=1}^N y_i = \gamma, \\
			      \hspace{0.5cm}  & y_i \ge 0, \quad i\in[1:N],
\end{align}
where $\gamma$ is a positive constant. The optimum solution is uniform, i.e., $y_1=\cdots=y_N=\frac{\gamma}{N}$.
\end{lemma}
\begin{IEEEproof}
Since $g(\cdot)$ is concave, $\sum_{i=1}^{N}g(y_i)$ is Schur-concave. Thus, using the fact that the uniform solution is majorized by all feasible solutions, the maximum is attained by the uniform solution.
\end{IEEEproof}

Now we are ready to prove Theorem \ref{thm:upper_IRR}.

\begin{IEEEproof}[Proof of Theorem \ref{thm:upper_IRR}]
Recall that $\kappa(\cdot)$ is a mapping from $[1:N]$ to $[1:M]$.
Denote $\mathcal{J}_i= \{\ell\in[1:N] \big| \kappa(\ell) = i\}$ for all $i\in[1:M]$. For all $\ell\in[1:N]$, let $P_\ell$ and $n_\ell$ denote the transmit power and the number of time slots used in round $\ell$, respectively. To satisfy the average power constraint, we must have 
\begin{align} \label{eq:ineq_P}
\sum_{\ell\in\mathcal{J}_i}n_\ell P_\ell \le nP 
\end{align}
for all $i\in[1:M]$, where $n=\sum_{\ell=1}^Nn_\ell$. For notational convenience, denote $r_\ell = I(S_{\kappa(\ell)};V^*_\ell|V^*_{[1:\ell-1]})$ for all $\ell\in[1:N]$.

For each $\ell\in[1:N]$, the $V^*_\ell$ can be decoded reliably as $k$ increases only if  
\begin{align}
\frac{k}{n_\ell} \le \frac{\frac{1}{2}\log\left(1+P_\ell\right)}{r_\ell}, 
\end{align}
and thus the achievable computation rate must satisfy
\begin{align}
R =\frac{k}{n} \le \min_{\ell\in[1:N]}\frac{n_\ell}{n} \frac{\frac{1}{2}\log\left(1+P_\ell\right)}{r_\ell}.
\end{align}
For convenience, define $\alpha_{i\ell} = \frac{n_\ell}{\sum_{\ell\in\mathcal{J}_i}n_\ell}$ and $\beta_{i} = \frac{\sum_{\ell\in\mathcal{J}_i}n_\ell}{n}$ for all $i\in[1:M]$, $\ell\in[1:N]$.
Then, the achievable computation rate can be upper bounded as 
\begin{align}
R & \le \min_{i\in[1:M]} \min_{\ell\in\mathcal{J}_i}\frac{n_\ell}{n} \frac{\frac{1}{2}\log\left(1+P_\ell\right)}{r_\ell} \\
& \overset{(a)}{\le} \frac{\sum_{i=1}^M\sum_{\ell\in\mathcal{J}_i}\frac{n_\ell}{n}\frac{1}{2}\log\left(1+P_\ell\right)}{\sum_{\ell=1}^Nr_\ell} \\
& = \frac{\sum_{i=1}^M\beta_{i}\sum_{\ell\in\mathcal{J}_i}\alpha_{i\ell}\frac{1}{2}\log\left(1+P_\ell\right)}{I(S_{[1:M]};V^*_{[1:N]})} \\
&\overset{(b)}{\le} \frac{\sum_{i=1}^M\beta_{i}\frac{1}{2}\log\left(1+\sum_{\ell\in\mathcal{J}_i}\alpha_{i\ell}P_\ell\right)}{I(S_{[1:M]};V^*_{[1:N]})} \\
&\overset{(c)}{\le} \frac{\sum_{i=1}^M\beta_{i}\frac{1}{2}\log\left(1+\frac{P}{\beta_i}\right)}{I(S_{[1:M]};V^*_{[1:N]})} \\
&\overset{(d)}{\le} \frac{\frac{1}{2}\log\left(1+MP\right)}{I(S_{[1:M]};V^*_{[1:N]})} 
\end{align}
where $(a)$ follows from Lemma \ref{lma:x}, (b) follows since the function $\log(1+x)$ with $x\ge 0$ is concave, (c) follows from \eqref{eq:ineq_P}, and $(d)$ follows from Lemma \ref{lma:y} and the fact that $x\log(1+P/x)$ is a concave function of $x$.
\end{IEEEproof}

\section*{Appendix B \\ Bounded Entropy of the Descriptions of the Clipped Frequencies as $M\to \infty$} 
In this appendix, we provide a proof of Lemma \ref{lma:suff_d} and then Lemma \ref{lma:suff_info} will follow as a special case with $d=0$. Since the proof works universally for every clipped frequency $\overline{b}_\ell$, we drop all indices $\ell$ in the proof for simplicity. Besides, with an abuse of notation, we denote $p_i := \P({S_i}=\ell)$ for all $i\in[1:M]$.
For the proof of Lemma \ref{lma:suff_d}, we need the following lemma, which upper bounds the entropy of the sum of independent Bernoulli random variables.
\begin{lemma}\label{lma:H_bsum}
Fix $p_{[1:M]} \in [0:1]^M$. Let $X_{[1:M]}$ be independent random variables, where $X_i\sim$ Bernoulli($p_i$) for all $i\in[1:M]$. Then, 
\begin{align}\label{eq:biH2_ine}
H\left(\sum_{i=1}^M X_i\right) \le \frac{1}{2} \log\left(2\pi e\left(\sum_{i=1}^Mp_i+\frac{1}{12}\right)\right).
\end{align}
\end{lemma}

\begin{IEEEproof}
First, applying Theorem $1$ in \cite{Shepp:78}, we have 
\begin{align}
\label{eq:non_uni}
H\left(\sum_{i=1}^MX_i\right) \le H\left(\sum_{i=1}^M\overline{X}_i\right),
\end{align} 
where $\overline{X}_{[1:M]}$ are i.i.d. Bernoulli$\left(\overline{p}\right)$ random variables and $\overline{p}=\frac{1}{M}\sum_{i=1}^Mp_i$. 

Notice that $\sum_{i=1}^M\overline{X}_i\sim$ Binomial$\left(M,\overline{p}\right)$. 
Let $Y$ be a Poisson random variable with mean $\sum_{i=1}^Mp_i$. Then, we have
\begin{align}
H\left(\sum_{i=1}^M\overline{X}_i\right) &\overset{(a)}{\le} H(Y) \\
\label{eq:prebiH2}
&\overset{(b)}{\le} \frac{1}{2}\log \left(2\pi e\left(\sum_{i\in\mathcal{A}_m}p_i+\frac{1}{12}\right)\right), 
\end{align}
where (a) follows from \cite[Theorems 7 and 8]{Harremoes:01} and (b) follows from \cite[Expression (1)]{Adell:10}. Finally, combining \eqref{eq:non_uni} and \eqref{eq:prebiH2}, the inequality \eqref{eq:biH2_ine} is established.
\end{IEEEproof}

\begin{IEEEproof}[Proof of Lemma \ref{lma:suff_d}]
Without loss of generality, we assume $\theta\le M$ since there are only $M$ sensors.  If $\theta=0$, then $U_m=0$ for all $m\in[1:J]$ and thus $H\left(U_{[1:J]}\right)=0$. In the following, we consider the case $1\le\theta\le M$. Since the sources are independent, $U_1\leftrightarrow\cdots\leftrightarrow U_{J}$ forms a Markov chain. From now on, we consider a fixed $d\in[0:J-1]$. 

First, consider the case $\sum_{i=1}^{M} p_i \le \theta$. In this case, we use the $M$-partition. Applying Lemma \ref{lma:H_bsum} by substituting $X_i$ with $\mathbf{1}_{\{S_i=\ell\}}$, we have 
\begin{align}
H(U_1) &= H\left(\sum_{i=1}^M\mathbf{1}_{\{S_i=\ell\}}\right) \\
&\le \frac{1}{2}\log \left(2\pi e\left(\sum_{i=1}^Mp_i+\frac{1}{12}\right)\right) \\
& \le \frac{1}{2}\log \left(2\pi e\left(\theta+\frac{1}{12}\right)\right).
\end{align}

Next, consider the case $\sum_{i=1}^{M} p_i > \theta$. Let the intervals $\left[a_{m-1}+1:a_m\right]$, $m\in[1:J]$, satisfy $0=a_0 < \cdots < a_{J}=M$, 
\begin{eqnarray}
\label{eq:set_rule}
 & \displaystyle\sum_{i=a_{m-1}+1}^{a_m-1}p_i < \theta \le\sum_{i=a_{m-1}+1}^{a_m}p_i, & \text{for } m\in[1:J-1], \\
\label{eq:set_rule2}
\text{and} & \displaystyle \theta \le \sum_{i=a_{J-1}+1}^{M}p_i < 2\theta. &
\end{eqnarray}
Note that for all $m\in[1:J-1]$, since $p_{a_m}\le 1$, \eqref{eq:set_rule} implies that 
\begin{align} \label{eq:set_imply}
\sum_{i=a_{m-1}+1}^{a_m}p_i < \theta + 1.
\end{align}
Set $\mathcal{A}_m = \left[a_{m(d)-1}+1:a_{m(d)}\right]$ for all $m\in[1:M]$, where $m(d) = ((m+d-1)\bmod J) +1$. 

Then, the entropy $H\left(U_{[1:J]}\right)$ can be upper bounded as follows.
\begin{align}
 H\left(U_{[1:J]}\right)\nonumber & =  \sum_{m=1}^{J} H(U_m|U_{[1:m-1]}) \\
&\overset{(a)}{=} H(U_1)+\sum_{m=2}^{J} H(U_m|U_{m-1}) \\
& \overset{(b)}{=} H(U_1)+ \sum_{m=2}^{J}  \sum_{j=0}^{\theta-1}\P\left(U_{m-1}=j\right)H\left(U_m\big|U_{m-1}=j\right) \\
\label{eq:HUbound}
& =H\left(\sum_{i\in\mathcal{A}_1}\mathbf{1}_{\{S_i=\ell\}}\right) + \sum_{m=2}^{J} \sum_{j=0}^{\theta-1} 
\P\left(U_{m-1}=j\right)H\left(\sum_{i\in\mathcal{A}_m}\mathbf{1}_{\{S_i=\ell\}}\right),
\end{align}
where {(a)} follows since $U_1\leftrightarrow\cdots\leftrightarrow U_{J}$ forms a Markov chain and (b) follows since $U_m$ conditioned on $\{U_{m-1}\ge\theta\}$ is deterministic. 

Then, Lemma \ref{lma:H_bsum}, \eqref{eq:set_rule2}, and \eqref{eq:set_imply} imply that if $m(d)\in[1:J-1]$, 
\begin{align}
H\left(\sum_{i\in\mathcal{A}_m}X_i\right) &< \frac{1}{2}\log \left(2\pi e\left(\theta+1+\frac{1}{12}\right)\right), 
\end{align}
and if $m(d)=J$, 
\begin{align}
\label{eq:biH2}
H\left(\sum_{i\in\mathcal{A}_m}X_i\right) &< \frac{1}{2}\log \left(2\pi e\left(2\theta+\frac{1}{12}\right)\right) < \frac{1}{2} + \frac{1}{2}\log \left(2\pi e\left(\theta+1+\frac{1}{12}\right)\right).
\end{align}
Hence, \eqref{eq:HUbound} to \eqref{eq:biH2} imply that  
\begin{align}
H\left(U_{[1:J]}\right) &<  \frac{1}{2} + \frac{1}{2}\log \left(2\pi e\left(\theta+\frac{13}{12}\right)\right) \left(1 + \sum_{m=2}^{J} \sum_{j=0}^{\theta-1} \P\left(U_{m-1}=j\right)\right) \\
\label{eq:HUbound2}
& <  \frac{1}{2} + \frac{1}{2}\log \left(2\pi e\left(\theta+\frac{13}{12}\right)\right) \left(4 + \sum_{m=5}^{J} \sum_{j=0}^{\theta-1} \P\left(U_{m-1}=j\right)\right).
\end{align}

Now we show that the double summation in \eqref{eq:HUbound2} can be upper bounded by a constant independent of $J$ and $\theta$. Denote $\mathcal{S}_0=\emptyset$ and $\mathcal{S}_{m}=\bigcup_{t=1}^{m}\mathcal{A}_t$ for all $m\in[1:J]$. For $j\in[1:|\mathcal{S}_{m-1}|]$, $\P\left(U_{m-1}=j\right)=\P\left(Y=j\right)$ where $Y\sim\text{Poisson Binomial}(p_{\mathcal{S}_{m-1}})$. Denote by $\mathcal{F}_m$ the set of all subsets of $\mathcal{S}_{m}$ with $j$ elements and let $\Omega_m^*\in \mathcal{F}_m$ be the set of the $j$ indices with the largest values of $p_i$. Then, we have 
\begin{align}
 \P\left(U_{m-1}=j\right) &= \sum_{\Omega\in\mathcal{F}_m} \prod_{i\in\Omega} p_i \prod_{t\in\mathcal{S}_{m-1}\backslash\Omega} \left(1-p_t\right) \\
& \le  \prod_{t\in\mathcal{S}_{m-1}\backslash\Omega_m^*} \left(1-p_t\right) \sum_{\Omega\in\mathcal{F}_m} \prod_{i\in\Omega} p_i \\
& \overset{(a)}{\le} \left(1-\frac{1}{\left|\mathcal{S}_{m-1}\right|-j}\sum_{t\in\mathcal{S}_{m-1}\backslash\Omega_m^*}p_t\right)^{\left|\mathcal{S}_{m-1}\right|-j} \sum_{\Omega\in\mathcal{F}_m} \prod_{i\in\Omega} p_i \\
& \overset{(b)}{\le} \left(1-\frac{(m-1)\theta-j}{\left|\mathcal{S}_{m-1}\right|-j}\right)^{\left|\mathcal{S}_{m-1}\right|-j} \sum_{\Omega\in\mathcal{F}_m} 
\prod_{i\in\Omega} p_i \\
& \overset{(c)}{\le} \left(e^{-1}\right)^{(m-1)\theta-j} \sum_{\Omega\in\mathcal{F}_m} 
\prod_{i\in\Omega} p_i \\
& \le  e^j e^{-(m-1)\theta} \frac{1}{j!}\left(\sum_{i\in\mathcal{S}_{m-1}} p_i\right)^j \\
\label{eq:biP}
& \overset{(d)}{\le} e^j e^{-(m-1)\theta} \frac{(m(\theta+1))^j}{j!},
\end{align}
where $(a)$ follows since $\prod_i x_i$ is Schur-concave when all $x_i > 0$, $(b)$ follows from \eqref{eq:set_rule} and \eqref{eq:set_rule2}, $(c)$ follows since $\left(1-\frac{u}{x}\right)^x \le e^{-u}$ for all $x\ge 1$, and $(d)$ follows from \eqref{eq:set_rule2} and \eqref{eq:set_imply}. Thus, 
\begin{align}
\sum_{m=5}^{J} \sum_{j=0}^{\theta-1} \P\left(U_{m-1}=j\right) &\le \sum_{m=5}^{J} \sum_{j=0}^{\theta-1} e^{-(m-1)\theta} \frac{(me(\theta+1))^j}{j!} \\
&= e^{-9\theta/4}\sum_{m=0}^{J-4} \sum_{j=0}^{\theta-1} \frac{((m+5)e(\theta+1))^j}{j!} e^{-(m+7/4)\theta} \\
&\overset{(a)}{\le} e^{-5\theta/4}\theta\frac{(\theta+1)^{\theta}}{\theta!} \sum_{m=0}^{J-4} (m+5)^{\theta}e^{-(m+7/4)\theta} \\
&\overset{(b)}{\le} e^{-5\theta/4}\theta\frac{(\theta+1)^{\theta}}{\sqrt{2\pi\theta}(\theta/e)^{\theta}} \sum_{m=0}^{J-4} (m+5)^{\theta}e^{-(m+7/4)\theta} \\
&=\frac{1}{\sqrt{2\pi}}e^{-\theta/4}\sqrt{\theta}\left(1+\frac{1}{\theta}\right)^{\theta} \sum_{m=0}^{J-4} \left((m+5)e^{-m-7/4}\right)^{\theta}  \displaybreak[4]\\
&\overset{(c)}{\le} \sqrt{\frac{e}{\pi}} \sum_{m=0}^{J-4} \left((m+5)e^{-m-7/4}\right)^{\theta} \\
\label{eq:series3m} 
&\overset{(d)}{\le} \sqrt{\frac{e}{\pi}}  \sum_{m=0}^{\infty} (m+5)e^{-m-7/4} <1 
\end{align}
where $(a)$ follows since $\frac{c^j}{j!}$ is an increasing function of $j$ for all $0\le j\le c$, (b) follows from Strling's formula, (c) follows since $\sqrt{m}e^{-m/4} \le \sqrt{2/e}$ for all $m\in\mathbb{Z}^+$ and $(1+1/x)^{x}<e$ for all $x>0$, and $(d)$ follows since $(m+5)e^{-m-7/4}<1$ for all $m\ge0$. Finally, we substitute \eqref{eq:series3m} into \eqref{eq:HUbound2} and then the theorem is established after some straightforward simplification.
\end{IEEEproof}

\bibliographystyle{IEEEtran}
\bibliography{IEEEabrv,References}
\end{document}